\documentclass[conference, 10pt]{IEEEtran}
\usepackage{amsfonts}
\usepackage{times}
\usepackage[pdftex]{graphicx}
\usepackage{latexsym}
\usepackage{dsfont}
\usepackage{amssymb}
\usepackage{amsmath}
\usepackage{cite}
\usepackage{verbatim}
\usepackage{subfigure}

\newcommand{\figref}[1]{{Fig.}~\ref{#1}}

\def\ba{{\mathbf{a}}}

\def\bv{{\mathbf{v}}}

\def\b0{{\mathbf{0}}}

\def\bA{{\mathbf{A}}}

\def\bH{{\mathbf{H}}}

\def\cA{\mathcal{A}}
\def\cB{\mathcal{B}}

\def\cF{\mathcal{F}}

\def\cL{\mathcal{L}}

\def\cP{\mathcal{P}}

\def\cS{\mathcal{S}}

\def\cW{\mathcal{W}}
\def\cX{\mathcal{X}}

\def\bell{{\mathbb{\ell}}}

\IEEEoverridecommandlockouts
\usepackage{cite}
\usepackage{amsmath,amssymb,amsfonts}
\usepackage{algorithmic}
\usepackage{graphicx}
\usepackage{textcomp}
\usepackage{xcolor}
\usepackage{soul}
\usepackage{float}

\usepackage{stfloats}
\usepackage{url}
\usepackage{verbatim}
\usepackage{algorithm}
\usepackage[top=0.71in,left=0.63in,right=0.63in,bottom=1.04in]{geometry}

\def\BibTeX{{\rm B\kern-.05em{\sc i\kern-.025em b}\kern-.08em
		T\kern-.1667em\lower.7ex\hbox{E}\kern-.125emX}}

\usepackage{amsmath}

\newcommand{\figsizeone}{0.50}
\newcommand{\figsizetwo}{0.42}
\newcommand{\figsizethree}{0.49}
\newcommand{\figsizefour}{0.48}
\newcommand{\figsizefive}{0.49}

\newcommand{\BS}{BS}
\newcommand{\UE}{UE}
\newcommand{\Hk}{\bH[k]}
\newcommand{\roi}[0]{\cX_\text{RoI}}

\newcommand{\fj}{\mathbf{f}_{j}}
\newcommand{\wi}{\mathbf{w}_{i}}

\newcommand{\yij}{y_{i,j}[k]}

\newcommand{\rate}{R}
\newcommand{\rateji}{\rate_{i,j}}
\newcommand{\ratejin}[3]{\rate_{#1,#2,#3}}

\newcommand{\Ndata}{N_\text{d}}
\newcommand{\beampair}{(\wi,\fj)}
\newcommand{\RTijn}[3]{R_{\text{T},#1,#2,#3}}
\newcommand{\RTin}[2]{R_{\text{T},#1,#2}}
\newcommand{\RTvec}{\textbf{\textit{r}}_{\text{T}}}
\newcommand{\RTvechat}{\hat{\textbf{\textit{r}}}_{\text{T}}}
\newcommand{\RTvecf}{\textbf{\textit{r}}_{\text{T},\text{f}}}
\newcommand{\RTvecw}{\textbf{\textit{r}}_{\text{T},\text{w}}}
\newcommand{\RTvecfhat}{\hat{\textbf{\textit{r}}}_{\text{T},\text{f}}}
\newcommand{\RTvecwhat}{\hat{\textbf{\textit{r}}}_{\text{T},\text{w}}}

\newcommand{\RTvecn}{\textbf{\textit{r}}_{\text{T},n}}
\newcommand{\RTvechatn}{\hat{\textbf{\textit{r}}}_{\text{T},n}}
\newcommand{\RTveckhatn}{\hat{\textbf{\textit{r}}}_{\text{T},\text{k},n}}
\newcommand{\RTveckn}{\textbf{\textit{r}}_{\text{T},\text{k},n}}
\newcommand{\RTvecfn}{\textbf{\textit{r}}_{\text{T},\text{f},n}}
\newcommand{\RTvecwn}{\textbf{\textit{r}}_{\text{T},\text{w},n}}

\newcommand{\RT}{R_\text{T}}

\newcommand{\belln}{\boldsymbol{\ell}_n}
\newcommand{\boldell}{\boldsymbol{\ell}}

\newcommand{\misalign}{P_{\text{m}}}

\newcommand{\card}[1]{|#1|}
\newcommand{\abs}[1]{|#1|}
\newcommand{\scenabbone}{CBSwL}
\newcommand{\scenabbtwo}{DBSwL}
\newcommand{\scenabbthree}{DBSwoL}

\begin{document}
\title{Beam Training in mmWave Vehicular Systems: Machine Learning for Decoupling Beam Selection \\ }
	
	\author{\IEEEauthorblockN{Ibrahim Kilinc\IEEEauthorrefmark{1},  Ryan M. Dreifuerst\IEEEauthorrefmark{2},  Junghoon Kim\IEEEauthorrefmark{3},     and Robert W. Heath Jr.\IEEEauthorrefmark{1}}
			\IEEEauthorblockA{\IEEEauthorrefmark{1}ECE Department, University of California, San Diego, USA
			(e-mail: \{ikilinc, rwheathjr\}@ucsd.edu)\\
            \IEEEauthorrefmark{2}ECE Department, NC State University, Raleigh, USA (e-mail: rmdreifu@ncsu.edu)\\
            \IEEEauthorrefmark{3}Motorola Mobility, Chicago, IL 60654, USA (email: junghoon@motorola.com)\\
		}
  }

\maketitle
\thispagestyle{empty}

%


\begin{abstract}

Codebook-based beam selection is one approach for configuring millimeter wave communication links. The overhead required to reconfigure the transmit and receive beam pair, though, increases in highly dynamic vehicular communication systems. Location information coupled with machine learning (ML) beam recommendation is one way to reduce the overhead of beam pair selection. 
In this paper, we develop ML-based location-aided approaches to decouple the beam selection between the user equipment (UE) and the base station (BS). We quantify the performance gaps due to decoupling beam selection and also disaggregating the UE's location information from the BS. Our simulation results show that decoupling beam selection with available location information at the BS performs comparable to joint beam pair selection at the BS. Moreover, decoupled beam selection without location closely approaches the performance of beam pair selection at the BS when sufficient beam pairs are swept.
\end{abstract}

\section{Introduction}\label{sec: introduction}

Millimeter wave (mmWave) multiple-input multiple-output (MIMO) communications promise enhanced connectivity with high-fidelity sensor data exchange in vehicular systems \cite{ChoiEtAlMillimeterWaveVehicularCommunicationSupport2016}. Obtaining the best performance in mmWave MIMO systems requires configuring transmit and receive antenna arrays, which is challenging with large arrays and hybrid architectures \cite{VaEtAlInverseMultipathFingerprintingMillimeter2018, Gonzalez-PrelcicEtAlMillimeterWaveCommunicationOutofBandInformation2017}. Codebook-based beam training is one approach for configuring mmWave MIMO links by transmitting and receiving with each beamforming codeword-pair in highly dynamic vehicular environments \cite{VaEtAlInverseMultipathFingerprintingMillimeter2018,Gonzalez-PrelcicEtAlMillimeterWaveCommunicationOutofBandInformation2017}. Prior work has demonstrated how sensory information from localization sensors \cite{WangEtAlMmWaveVehicularBeamSelection2019,Reus-MunsEtAlDeepLearningVisualLocation2021,ZecchinEtAlLIDARPositionAidedMmWaveBeam2022}, camera \cite{Reus-MunsEtAlDeepLearningVisualLocation2021}, LiDAR \cite{ZecchinEtAlLIDARPositionAidedMmWaveBeam2022}, radar \cite{yunchenradar} can be leveraged to reduce the overheads of beam training. Most prior work on ML-based beam training focuses on a centralized approach where the BS leverages sensor data to recommend beam pairs to test \cite{WangEtAlMmWaveVehicularBeamSelection2019,Reus-MunsEtAlDeepLearningVisualLocation2021,ZecchinEtAlLIDARPositionAidedMmWaveBeam2022, yunchenradar}. This requires the recommended receive beams to be shared with the UEs and requires all UEs to have the same codebook and antenna configuration.

In this work, we develop location-aided beam training approaches that decouple the beam selection at the BS and the UE. We consider three scenarios. \textit{Scenario 1} represents our baseline case where the BS determines the beam pairs for the BS and the UE based on the UE's location information. In the \textit{scenario 2}, the BS selects its transmit beams based on the UE location. Independently, the UE determines its receive beams by leveraging its location. In the \textit{scenario 3}, the BS does not have the information of the UE's location and the selected transmit beams are chosen to serve the region of interest in the urban street. The UE selects its beam based on its location information. We denote beam selection in the \textit{scenario 1} as \textit{coupled with location}, and beam selection in the \textit{scenario 2} and \textit{3} as \textit{decoupled with} and \textit{without location}. We develop ML-based beam selection algorithms for coupled and decoupled scenarios because ML has been shown to successfully learn implicit relationships between beams and location information in site-specific scenarios \cite{dreifuerst2024neural}. The algorithms for \textit{scenario 2} and \textit{3} are based on lightweight ML models for practical deployment at the UE. We generate ray-traced channel samples in a realistic urban environment. Accordingly, we compare the three scenarios to quantify the performance gaps due to decoupling beam selection and disaggregating the UE’s location from the BS.

There are various approaches for side information-aided, ML-powered beam selection for vehicular mmWave systems. Leveraging vehicle location in beam pair selection has been proposed in \cite{WangEtAlMmWaveVehicularBeamSelection2019,VaEtAlInverseMultipathFingerprintingMillimeter2018,SatyanarayanaEtAlDeepLearningAidedFingerprintBased2019}. The work shows a decrease in the overhead, whereas, the algorithms are for coupled beam selection and are centralized at the BS. The images taken by a roadside unit (RSU) and the UE's location information were fused to develop a beam recommendation algorithm in \cite{Reus-MunsEtAlDeepLearningVisualLocation2021}. The beam selection at the BS is again coupled. Similar to coupled beam selection methods, LiDAR and location measurements were processed to predict beam pairs for the BS and the UE in \cite{ZecchinEtAlLIDARPositionAidedMmWaveBeam2022}. Compared to \cite{WangEtAlMmWaveVehicularBeamSelection2019,VaEtAlInverseMultipathFingerprintingMillimeter2018,SatyanarayanaEtAlDeepLearningAidedFingerprintBased2019, Reus-MunsEtAlDeepLearningVisualLocation2021,ZecchinEtAlLIDARPositionAidedMmWaveBeam2022}, we propose ML-based algorithms decoupling the beam selection at the BS and the UE, which might be a practical approach to incorporate different UE configurations into beam training process. Unlike \cite{ZecchinEtAlLIDARPositionAidedMmWaveBeam2022,Reus-MunsEtAlDeepLearningVisualLocation2021}, we only focus on the UE's location information as it is among the simplest sensory information to obtain and share.

\noindent
\textbf{Notation}: $\bA$ is a matrix, $\ba$ is a column vector, $\cA$ is a set and a, A denote scalars. $(\cdot)^*$ is conjugate transpose, $\card{\cA}$ is the cardinality of set $\cA$, $\mathds{1}(\cdot)$ is the indicator function and $\abs{\text{a}}$ is the absolute value of a scalar a. $\times$ denotes the Cartesian product of two sets, $\subset$ denotes the subset symbol in sets. 

\section{System model and problem formulation}\label{sec: system_model}
In this section, we formally introduce the system model, the general beam training problem, and the location-based dataset to help solve the problem.

\subsection{System model}

We consider a vehicular communication system in an urban street. We assume a BS at the roadside of a multi-lane road and a communicating vehicular UE on the road. The BS and the UE are equipped with uniform planar arrays (UPAs) for analog beamforming. We model a frequency-selective MIMO orthogonal frequency division multiplexing (OFDM) multipath channel with $K$ subcarriers. We denote the channel at the subcarrier $k$ as $\Hk$, and the BS and UE beam codebooks as $\cF$ and $\cW$. We denote the $j$-th beamformer at the {\BS} as $\fj \in \cF$, the $i$-th combiner at the {\UE} as $\wi \in \cW$ and the noise vector as $\bv$.
The received signal at the UE on the subchannel $k$ with the beam pair $\beampair$ is written as
\begin{equation}\label{eqn: received signal}
    \yij=  \wi^*\Hk\fj s[k] + \wi^*\bv[k].
\end{equation}
We denote the expected noise power over all subcarriers as $\sigma^2$ and assume beamformers and combiners have unit power. The average rate $\rateji$ obtained at the UE with $\beampair$ over $K$ subcarriers is calculated as

\begin{equation}\label{eqn: P_n_ij}
     \rateji = \frac{1}{K} \sum\limits_{k=0}^{K-1}{\log_2\left(1+{\dfrac{\abs{\yij}^2-\sigma^2}{\sigma^2}}\right)}.
\end{equation}
The average rate for beam pairs is used for beam selection.

Now, we explain the general beam selection problem for any scenario. Let $\cB$ denote the set of all beam pair combinations $\cW \times \cF$ and $\cS \subset \cB$ denote the beam pair subset recommended by a beam selection algorithm for the BS and the UE, then the throughput ratio \cite{WangEtAlMmWaveVehicularBeamSelection2019} is defined as 
\begin{equation}\label{eqn: throughput ratio n-th}
        R_{\text{T}} =\dfrac{\max_{\beampair \in \cS}{\rateji}}{\max_{\beampair\in \cB}{\rateji}}.
\end{equation}
We focus on a specific region inside the coverage area of the BS in the urban street. We denote the set of locations that a UE might be located as $\cX_\text{RoI}$. The region of interest can be defined arbitrarily in the coverage region and the problem formulation is not specific to the region, although the specific results and algorithms are site-specific. The urban street has both mobile and static objects. Consequently, the multipath channel between the BS and a UE is location-dependent and time-varying. Given a realization of channel for a UE, the throughput ratio depends only on $\cS$. We define the general beam selection problem as generating an $\cS$ for a given UE at $\boldell \in \cX_\text{RoI}$ such that
\begin{equation}\label{eqn:general_optim}
    \underset{\cS}{\text{argmax}}\text{ }{R_{\text{T}}}, \text{ subject to } \card{\cS} = N_\text{B}.
\end{equation}
The main goal in each scenario is to construct an $\cS$ to maximize the throughput ratio for a UE that might be located at any $\boldell \in \cX_\text{RoI}$. 

We develop a beam selection method for each scenario to generate a beam pair subset $\cS$. The task of generating $\cS$ involves three stages: \textit{training}, \textit{inference}, and \textit{beam-sweeping}. A location-based dataset is constructed in the \textit{training} stage. The beam selection methods are configured using the dataset to achieve \eqref{eqn:general_optim}. In the \textit{inference} stage, a subset of beam pairs $\cS$ for the BS and the UE in the urban street is generated through configured beam selection methods depending on the scenario. In the \textit{beam-sweeping} stage, each beam pair $\beampair \in \cS$ is tested between the BS and the UE so that the pair providing the greatest throughput ratio is found.

\subsection{Beam training dataset}

Exhaustive measurements for all beam pairs mapped with the UE location can be collected to construct a beam training dataset. The UEs measure their location via GPS or localization sensors \cite{ChoiEtAlMillimeterWaveVehicularCommunicationSupport2016}. The rate for each beam pair $\beampair \in \cB$ for $\Ndata$ UEs constitutes a location-based dataset to configure the methods for the beam selection. We assume that the location-based dataset illustrated in Table \ref{table: dataset} is available at the BS and the UE to develop beam selection methods for all scenarios in the training stage. The location information is encoded as a vector $\boldell = [x,y]^T$ in Cartesian coordinates, where the BS is the origin and the road is in the $y$-axis. We assume that UEs' heights are the same so we only consider $x,y$.

\begin{table}
\normalsize
\centering
\caption{Location-based dataset. The $k$-th row corresponds to the rates for beam pairs $\beampair$ of a ue at  $[x_k,y_k]$.}
\begin{tabular}{ c c c c }
    \hline
    \hline
    \text{Location} & $i,j =1,1$ & $i,j$ & $i,j=\card{\cW},\card{\cF}$\\
    \hline
    $\boldell_\text{1}$ & $\ratejin{1}{1}{1}$ & $\ratejin{i}{j}{1}$ & $\ratejin{\card{\cW}}{\card{\cF}}{1}$  \\
    ... &... &... &...\\
    $\boldell_{\Ndata}$ & $\ratejin{1}{1}{\Ndata}$ & $\ratejin{i}{j}{\Ndata}$ & $\ratejin{\card{\cW}}{\card{\cF}}{\Ndata}$ \\
    \hline
    \hline
\end{tabular}
\label{table: dataset}
\end{table}

\section{Beam alignment scenarios} \label{sec: scenarios_problem_formulation}

In this section, we explain the training, inference, and beam-sweeping stages for each of the three scenarios. There are two key factors that distinguish the scenarios. The first factor is the availability of UE's location information at the BS during the inference stage. The second factor is the type of beam selection: coupled or decoupled. In all scenarios, UEs are assumed to have their location information through their localization sensors. 

We achieve beam selection with location by throughput ratio prediction. Therefore, we transform the dataset of rates in Table \ref{table: dataset} into the dataset of throughput ratios to be used in all scenarios. Denote the maximum rate for the $n$-th row in Table \ref{table: dataset} as $R_{\text{max},n}$, then throughput ratio for $\beampair$ is calculated as $R_{\text{T},i,j,n} = R_{i,j,n}/R_{\text{max},n}$. We define the throughput ratio vector for the $n$-th row as $\textbf{\textit{r}}_{\text{T},n} = [\RTijn{1}{1}{n},...,\RTijn{\card{\cW}}{\card{\cF}}{n}]^{\text{T}}$, and construct the throughput ratio dataset.

\subsection{Scenario 1}
We first consider a centralized scenario where the BS determines the BS and UE beams using the UE's location. The BS shares the corresponding UE beams with the UE before the beam-sweeping stage. This scenario is called \textit{coupled} beam selection with location (\scenabbone). The problem is to select a beam pair set $\cS \subset \cB$ at the BS to optimize \eqref{eqn:general_optim}. Given the location $\boldell$ of the UE, selecting the beam pairs having $N_\text{B}$ greatest throughput ratios is one possible approach to solve the problem. We need to predict the throughput ratios for beam pairs for a specific UE location $\boldell \in \roi$.

\begin{figure}
    \centerline{ \includegraphics[width = \figsizeone \textwidth]{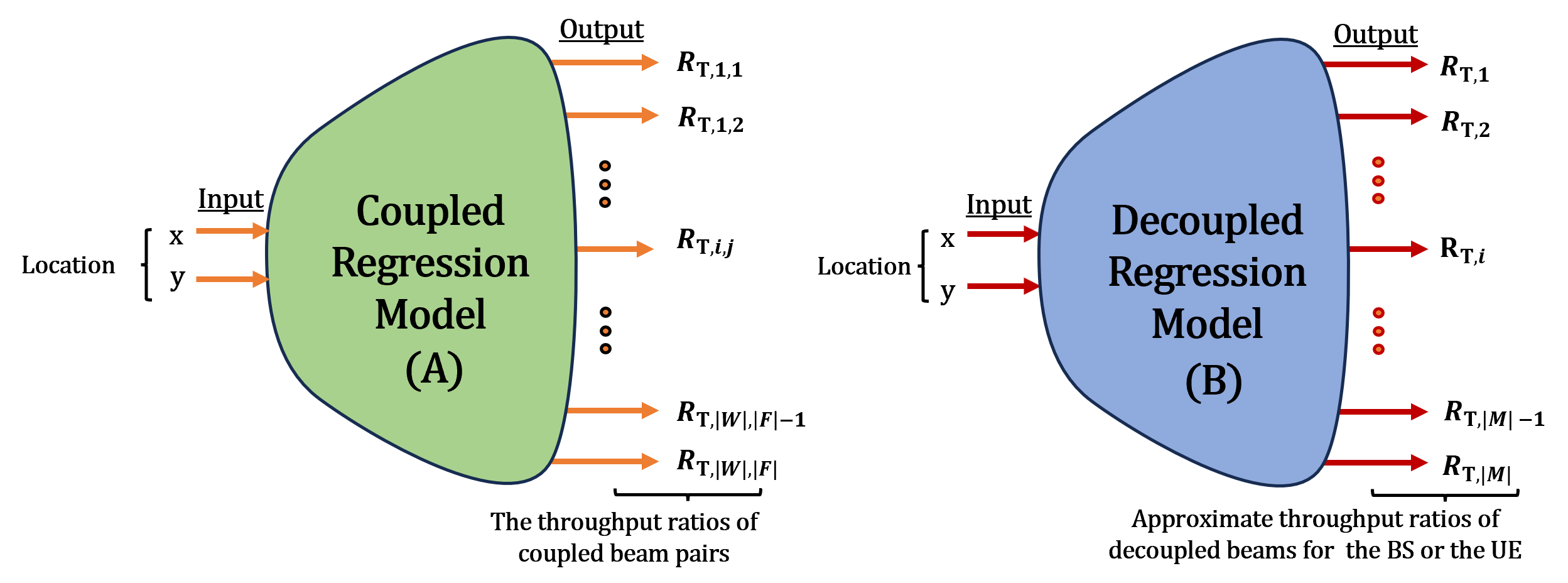}
 }
    \caption{Multi-output throughput ratio regression models are shown. Model (A) represents the regression model for \textit{scenario 1}. Model (B) represents the regression model for the BS or a UE in the \textit{scenario 2} and a UE in the \textit{scenario 3}, where \textit{M} denotes the beam codebooks, $\cF$ or $\cW$.}
    \label{fig:selection_models}
\end{figure}

In the training stage, we develop a regression model $g$ predicting $\RTvec$ given $\boldell$, illustrated in \figref{fig:selection_models}-(A). Let $\mathbf{\Theta}_1$ denote the trainable model parameters, $\RTvechat$ denote the predicted throughput ratio vector, then $\RTvechat=g(\bell|\mathbf{\Theta}_1)$. We use ML approaches to configure model parameters using the throughput ratio dataset to minimize the sum of prediction loss function $\cL$ over data points in Table \ref{table: dataset}
\begin{equation}
    \underset{\mathbf{\Theta}_1}{\text{min}} \sum_{n=1}^{N_d}{\cL(\RTvecn,\RTvechatn|\mathbf{\Theta}_1,\belln)}.
\end{equation}
The regression model $g$ incorporates the dynamic environment given the location $\boldell$ into the throughput ratio prediction. 

In the inference stage, a UE sends its location information $\boldell$ with a beam training request to the BS. The trained regression model predicts the throughput ratios using the location as $\RTvechat = g(\boldell|\mathbf{\Theta}_1)$. The selection approach is to construct $\cS$ containing beam pairs with the greatest $N_\text{B}$ throughput ratios. In this stage, the BS sends beam indices to the UE, requiring a transmission overhead of $\card{\cS}\log_2(\card{\cW})$ bits. Afterward, the beam-sweeping stage follows.

\subsection{Scenario 2}
Next we consider a decentralized scenario where each of the BS and UE determines its own beams given the UE's location. This scenario is called a \textit{decoupled} beam selection with location (\scenabbtwo). In the decoupled scenario, the BS and the UE select beam subset $\cS_\text{f} \subset \cF$ and $\cS_\text{w} \subset \cW$ both based on $\boldell$ to optimize \eqref{eqn:general_optim} where $\cS = \cS_\text{w} \times \cS_\text{f}$. Decoupled beam selection terminates the joint selection of beam pairs given the location. Accordingly, the UE and the BS cannot precisely predict the beams in $\cS_\text{f}$ and $\cS_\text{w}$.  Therefore, each terminal needs to select its combiners or beamformers based on the information available.

The beam pair with the highest throughput ratio might not necessarily lead to a reliable link all the time due to the dynamic environment. Therefore, selecting beamformers and combiners that yield the greatest throughput ratios may not lead to a high throughput ratio at the same location over time. Except for the targeted individually pairwise beam pairs, the other combinations of beamformers and combiners lead to bad combinations. Instead, we introduce the approximate throughput ratio (ATR) for each beam, aiming to provide resilience to the dynamic environment while achieving a high throughput ratio.
Given the $n$-th row, the ATR of the combiner $\wi$ is calculated by taking the average of throughput ratios for $\beampair$ over every $\fj \in \cF$. 
Likewise, the ATR of the beamformer $\fj$ is calculated over all combiners $\wi \in \cB$. Via this procedure, ATR datasets for $\cF$ and $\cB$ are generated. The dataset of the BS and the UE consists of ATR vectors, $\RTvecfn = [\RTin{1}{n},...,\RTin{\card{\cF}}{n}]$ and $\RTvecwn = [\RTin{1}{n},...,\RTin{\card{\cW}}{n}]$ for the $n$-th row. Two datasets represent the approximate achievable throughput ratio of beamformers and combiners. Our proposed solution to the decoupled problem is that the BS and the UE select beamformers and combiners with maximum ATRs.

In the training stage, we develop decoupled regression models, $g_\text{f}$ for the BS and $g_\text{w}$ for the UE. The regression models, $g_\text{f}$ and $g_\text{w}$ predict $\RTvecf$ and $\RTvecw$ given $\boldell$ for the UE, illustrated in \figref{fig:selection_models}-(B). Let $\mathbf{\Theta}_{2,\text{f}}$ $\mathbf{\Theta}_{2,\text{w}}$ be the trainable model parameters, $\RTvecfhat$ and $\RTvecwhat$ be the predicted ATR vectors, then $\RTvecfhat=g_\text{f}(\boldell|\mathbf{\Theta}_{2,\text{f}})$ and $\RTvecwhat=g_\text{w}(\boldell|\mathbf{\Theta}_{2,\text{w}})$. We train the regression models using the ATR datasets to minimize the loss between the true ratios and predicted ratios $\cL$
\begin{equation}\label{eqn: loss_scenario2}
    \underset{\mathbf{\Theta}_{2,k}}{\text{min}} \sum_{n=1}^{N_\text{d}}{\cL(\RTveckn,\RTveckhatn|\mathbf{\Theta}_{2,k},\belln)}, \text{for } k\in\{\text{f},\text{w}\}.
\end{equation}
Since the location-based dataset contains time-varying measurements, the regression models $g_\text{f}$, $g_\text{w}$ incorporate the effect of dynamic environment into ATR prediction given a location.

In the inference stage, a beam training request with the location information $\boldell$ is sent from a UE to the BS. $\cS_\text{f}$ and $\cS_\text{w}$ containing beams with the greatest ATRs in $\RTvecfhat=g_\text{f}(\boldell|\mathbf{\Theta}_{2,\text{f}})$ and $\RTvecwhat=g_\text{w}(\boldell|\mathbf{\Theta}_{2,\text{w}})$ are selected such that $\card{\cS}=\card{\cS_\text{w}\times\cS_\text{f}}=N_\text{B}$. $\card{\cS_\text{f}}$ and $\card{\cS_\text{w}}$ can take different values to satisfy $\card{\cS_\text{w}}\card{\cS_\text{f}}=N_\text{B}$. 
In this scenario, there is no transmission overhead from the BS to the UE required in the \textit{scenario 1}, since the combiners are selected at the UE. After beam selection at the BS and the UE, beam-sweeping starts.

\subsection{Scenario 3}
We consider a decentralized scenario where each of the BS and UE determines its own beams, but the BS does not have the UE's location in the inference stage, while the UE has its location information. We call this scenario \textit{decoupled} selection without location (\scenabbthree). Since the BS does not have the location information of a UE to serve, a reasonable strategy for the BS is to select beams, $\cS_\text{f} \subset \cF$ to cover the entire region that the UE might be located. The UE needs to select $\cS_\text{w} \subset \cW$ based on its location $\boldell$. The optimization problem in \eqref{eqn:general_optim} is to construct a beam pair set as $\card{\cS}= \card{\cS_\text{w}\times\cS_\text{f}}=N_\text{B}$. We propose selecting $\cS_\text{w}$ as in the \textit{scenario} 2 by maximizing ATR for a specific location $\boldell \in \roi$. The beamformer subset $\cS_\text{f}$ is selected to cover the region of interest in the urban street.

In the training stage, a decoupled regression model for a UE is developed to predict ATRs for $\cW$, $\RTvecwhat = g_\text{w}(\boldell|\mathbf{\Theta}_{3,\text{w}})$ as in the \textit{scenario} 2. 
Using the model, the beams providing the greatest ATRs are selected for $\cS_\text{w}$ for the UE. The BS beam selection cannot be conducted for a single UE since the BS does not have the UE's location. The multipath channel characteristics of a small region in the urban street might be highly correlated over time due to static objects, reflectors in the environment such as buildings and roads \cite{VaEtAlInverseMultipathFingerprintingMillimeter2018}. We incorporate this idea by proposing a clustering approach to group the measurements in the location-based ATR dataset via the K-means clustering algorithm \cite{kmeans}.  K-means clustering helps to create smaller grids in $\roi$. We design a probabilistic beam selection from each grid to cover the region of interest. Since the probabilities depend on multiple measurements in a grid, the approach provides robustness in a dynamic environment.

The training stage consists of clustering and the probability assignments on the beams. Initially, we divide beamformers' ATR dataset into $C$ clusters based on the location information. We introduce a parameter $\alpha$ to represent the significance of the information a cluster provides. Denoting the number of UEs in the cluster $c$ as $N_\text{c}$, the significance of cluster $c$ is defined as $\alpha_c = N_\text{c}/\Ndata$, where $\Ndata$ is the total number of measurements in the dataset. This definition assumes that the location-based dataset inherently represents a realistic UE distribution in the urban street. Otherwise, $\alpha$ for each cluster can be set to $1$ to discard the UE distribution in the dataset. We introduce a probability measure to compare the beamformers within and across the clusters based on the significance of each cluster. Denote the index of the beam with $k$-th highest ATR for the $n$-th row of the dataset in the cluster $c$ as $\{R_{\text{T},c,n}\}^k$, then the probability of being $k$-th best beam for $\fj$ is calculated as
\begin{equation}\label{eqn: k-th-best}
    P_c^{(k)}(j) = \dfrac{1}{N_\text{c}}\sum\limits_{n=1}^{N_\text{c}}{\mathds{1}\left(\{R_{\text{T},c,n}\}^k == j\right)}.    
\end{equation}
$P_c^{(k)}(j)$ incorporates the information of $k$-th strongest path for the cluster $c$ region. 

The beam selection from clusters starts with calculating $P_c^{(k)}(j)$ for each beamformer $\fj \in \cF$ in the clusters. The beams having non-zero probabilities are stored in the set $\cP_{c,k}$ for each cluster $c$. The $\ell$-th most probable beam of $\cP_{c,k}$ is denoted as $\{\cP_{c,k}\}^\ell$, where $\ell=1,...,\card{\cP_{c,k}}$. The $\ell$-th most probable beams are selected from each cluster. Some selected beams may yield high ATRs for other clusters, while some are good only within their clusters. Therefore, we take weighted averages of the probabilities of $\ell$-th most probable beams using the significance value of each cluster. This operation measures the generalizability of a beamformer in other regions. The idea is to select the most general beams across the clusters. The beam selection starts with $k = 1, \ell=1$. The beams are iteratively added to the $\cS_\text{f}$ for $\ell=1,...,|\cP_{c,k}|$ and for each $k=1,...,\card{\cF}$ until $\card{\cS_\text{f}}=N_\text{BS}$. This procedure converges quickly for the desired value of $N_\text{BS}$ beams. 

In the inference stage, a UE sends a beam training request to the BS. The BS has its selected beams ready in the training stage for beam-sweeping since the BS beams are irrespective of the UE's location. The UE selects $\cS_\text{w}$ based on the predetermined $\card{\cS_\text{w}}$ and $\boldell$ using $g_\text{w}(\boldell|\mathbf{\Theta}_{3,\text{w}})$. There is no need for any information exchange in this scenario.

\section{Simulation results} \label{sec: simulation_results}
We first introduce evaluation metrics and the simulation environment to generate the beam training dataset. Then we explain how learning models are trained and selected. Lastly, we present the performance comparison of the three scenarios.
\subsection{Evaluation metrics}
We use throughput ratio and misalignment probability as evaluation metrics in our simulations. Let $N$ denote the total number of test UEs selected from different snapshots of the urban street in \figref{fig:urban_street}. The performance metrics are calculated over all test UEs. Misalignment probability $\misalign$ is the probability that the beam pair subset does not include the beam pair with the greatest throughput ratio \cite{VaEtAlInverseMultipathFingerprintingMillimeter2018}. Let $\cS_{n}$ be the constructed beam pair subset for $n$-th test UE, it is defined as
\begin{equation}\label{eqn: metric_misalign}
    \misalign = \dfrac{1}{N}\sum\limits_{n = 1}^{N}{\mathds{1}\left(\underset{\beampair\in \cB}{\text{argmax}}{\{\RTvecn\}}\notin \cS_n\right)}.
\end{equation}
The average throughput ratio over $N$ test UEs is expressed as 
\begin{equation}\label{eqn: metric_througput}
    \RT = \dfrac{1}{N}\sum\limits_{n = 1}^{N}{\underset{\beampair\in \cS_n}{\text{max}}{\{\RTvecn\}}}.
\end{equation} 
Metrics highlight different performances of beam selection.

\subsection{Simulation setup}
\begin{figure}
    \centerline{ \includegraphics[width=\figsizetwo \textwidth]{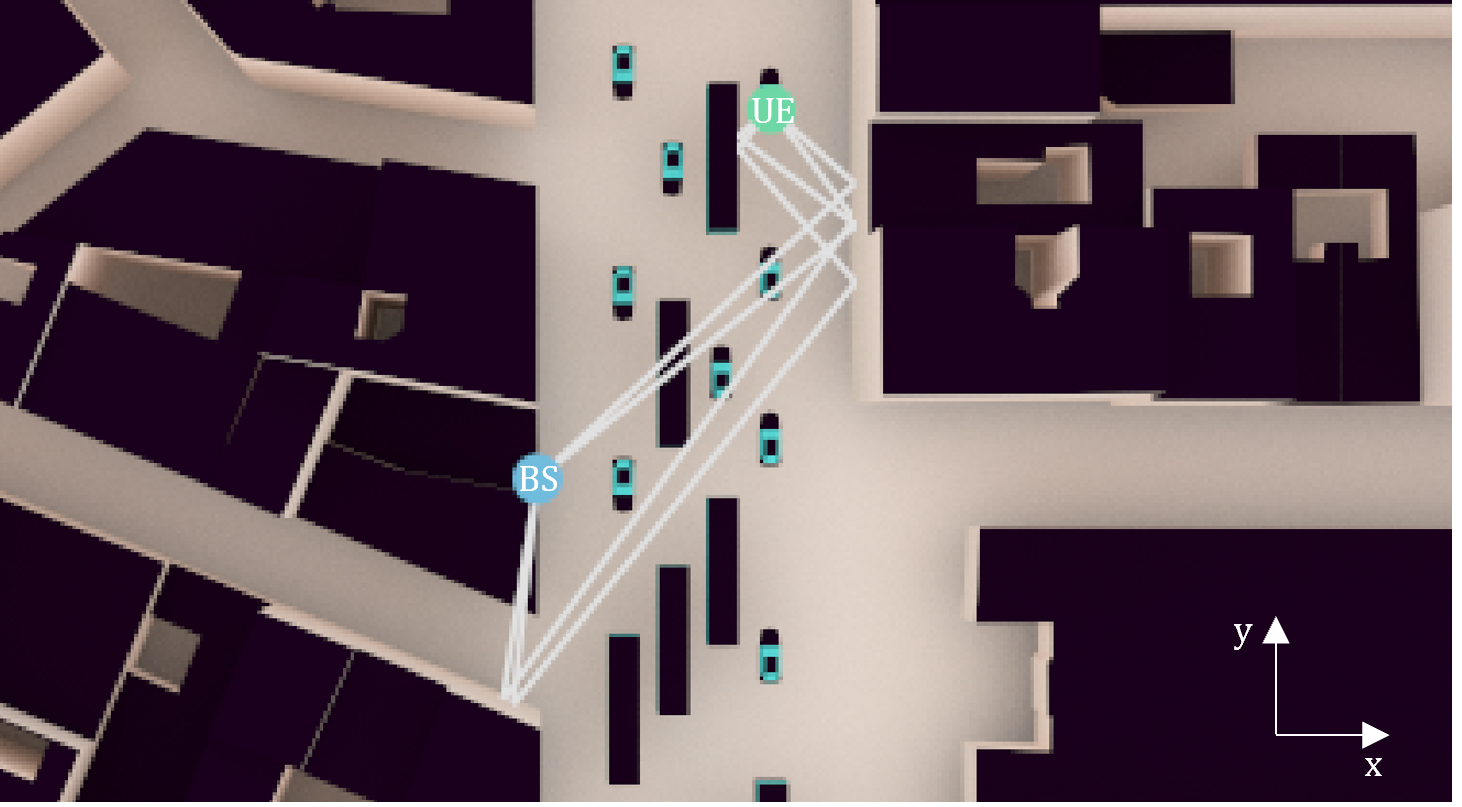}}
    \caption{One snapshot of the urban street environment created in Blender and simulated in Sionna. It represents the region of interest in the urban street. The buses are blockers and support NLOS paths from the BS and UEs.}
    \label{fig:urban_street}
\end{figure}

In our simulations, we use realistic channels generated from a ray tracing simulator, Sionna \cite{hoydis2023sionna} in an urban street environment. 
We use Blender to import realistic urban streets with buildings from OpenStreetMap \cite{OpenStreetMap}. We design cars (width: $1.75$m, length: $4.5$m, height: $1.5$m) and buses ($2.5$m, $12$m, $3.8$m) with realistic shapes in Blender. The walls of buildings and roads are made of concrete, the bodies of vehicles are made of metal, and the windows of vehicles are made of glass. The environment in the \figref{fig:urban_street} consists of a four-lane road where vehicles are placed with distances drawn from a uniform distribution. Cars serve as users, while the buses serve as natural reflectors and 
blockers in the environment.

We generate $500$ snapshots with different deployments of cars and buses in the urban street. \figref{fig:urban_street} shows an example snapshot and associated paths between the BS and a UE. The carrier frequency is $28$ GHz. We assume zero-mean additive Gaussian noise for the thermal noise and the noise figure in the receiver. An $8 \times 8$ UPA is mounted on the wall of a building, positioned $10$ meters above ground, serving as the BS and is tilted down so that boresight points in the street direction. A $4 \times 4$ UPA is placed on the roof of each car with a fixed orientation. We use DFT codebooks with the sizes $\card{\cF} = 64$ at the BS and $\card{\cW} = 16$ at the UE. The total number of beam pairs in the dataset is $\card{\cB}=\card{\cW}\card{\cF} = 1024$. We generate $5700$ channels for all cars in $500$ snapshots in the Sionna. 

\subsection{Model selection and training}\label{sec:stat_sig_models_train}

We design the beam selection algorithm by using ML, specifically, multi-output throughput ratio regression models for all scenarios except the decoupled
beam selection method of the BS in the \textit{scenario 3}. The fact that UE has limited storage and can employ only a lightweight ML model requires considering
the practical aspects of decoupled beam selection. Therefore, the UE’s decoupled regression model should be lightweight yet maintain satisfactory prediction performance. We compare XGBoost, neural network, random forest, and lasso linear regression models in the \textit{scenario 1} to identify the best-performing model with minimal parameters. We train and tune the models using K-fold cross-validation \cite{kfold}, an effective method to interpret the model's generalizability.

We split the simulated $5700$ channel samples into $10$ folds. The channels of $9$ folds are for constructing the location-based dataset and the remaining channels are for testing the model performances. The average model performances are for fine-tuning the hyperparameters. We maintain a similar number of trainable parameters across different learning models to ensure a fair comparison. XGBoost outperforms the other three in terms of misalignment probability and throughput ratio. Therefore, we use the XGBoost model for throughput ratio regression for all scenarios. The total number of trainable parameters in the decoupled regression model at the UE is kept significantly lower than the coupled regression model at the BS to consider practical aspects. The UE models of \textit{scenario 2} and \textit{3} have a number of parameters less than $2\times\card{\cB}$, whereas the XGBoost model in the \textit{scenario 1} has $30\times$ more parameters. The models are trained and tuned via K-fold cross-validation for $80\%$ of the generated channel samples. The remaining $20\%$ of channels are used as test samples for simulation results.

\subsection{The comparison of scenarios}
\begin{figure}
	\centering
	
	\includegraphics[width= \figsizethree \textwidth,trim={0 0 0 38},clip]{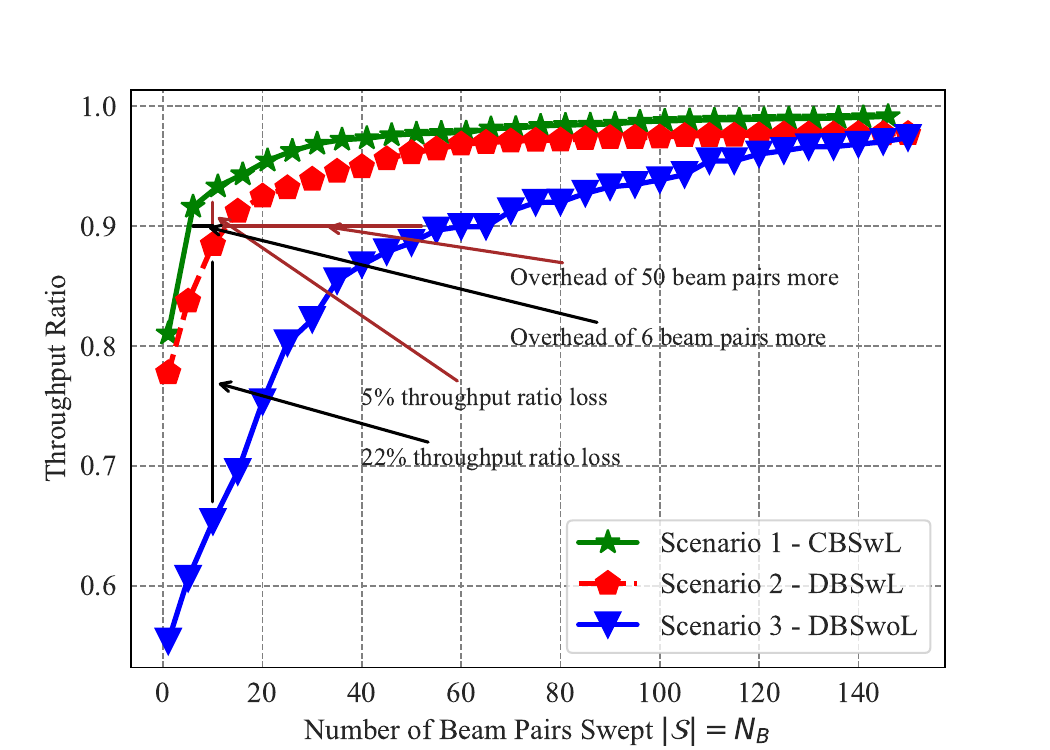}

	\caption{Throughput ratio comparison for the three scenarios. The number of UE beams, $\card{\cS_\text{w}}$ is set to 5 for the decoupled scenarios. The increasing number of beam pairs is due to the increasing number of selected BS beams. Decoupling beam selection with location has a minor throughput ratio decrease of less than $5\%$. Disaggregating location information has a higher decrease, whereas the proposed solutions achieve a comparable throughput ratio in sweeping at least $100$ beam pairs that are $\approx 10\%$ of total beam pairs.
	}
	\label{fig:throughput_ratio_all}
\end{figure} 

In this subsection, we compare the performance of the three scenarios in terms of throughput ratio and misalignment probability. \figref{fig:throughput_ratio_all} illustrates the throughput ratio comparison of the three scenarios. Our baseline in the \textit{scenario 1} performs quite well, achieving above $90\%$ throughput ratio in sweeping only $5$ beam pairs. Decoupling beam selection with location performs similarly to the baseline case with a throughput ratio decrease of less than $5\%$ for any number of beam pairs. This shows that the proposed transformation from throughput ratios to ATRs preserves the information of beam pairs achieving high throughput ratios. Moreover, sweeping only one beam pair achieves approximately $80\%$ throughput ratio in the beam selection with location. This highlights how impactful the location information is. In parallel, disaggregating the UE location from the BS in the \textit{scenario 3} leads to up to $22\%$ throughput ratio decrease in sweeping less than $100$ beam pairs. This also highlights the importance of the UE's location in the beam selection. Nevertheless, the proposed clustering-based beam selection algorithm in the \textit{scenario 3} incorporates the UE's location in the dataset into the beam selection well and quickly recovers the throughput ratio loss up to selecting $120$ beam pairs.

\begin{figure}[H]
	\centerline{
		\includegraphics[width=\figsizefour \textwidth,trim={50 0 50 23},clip]{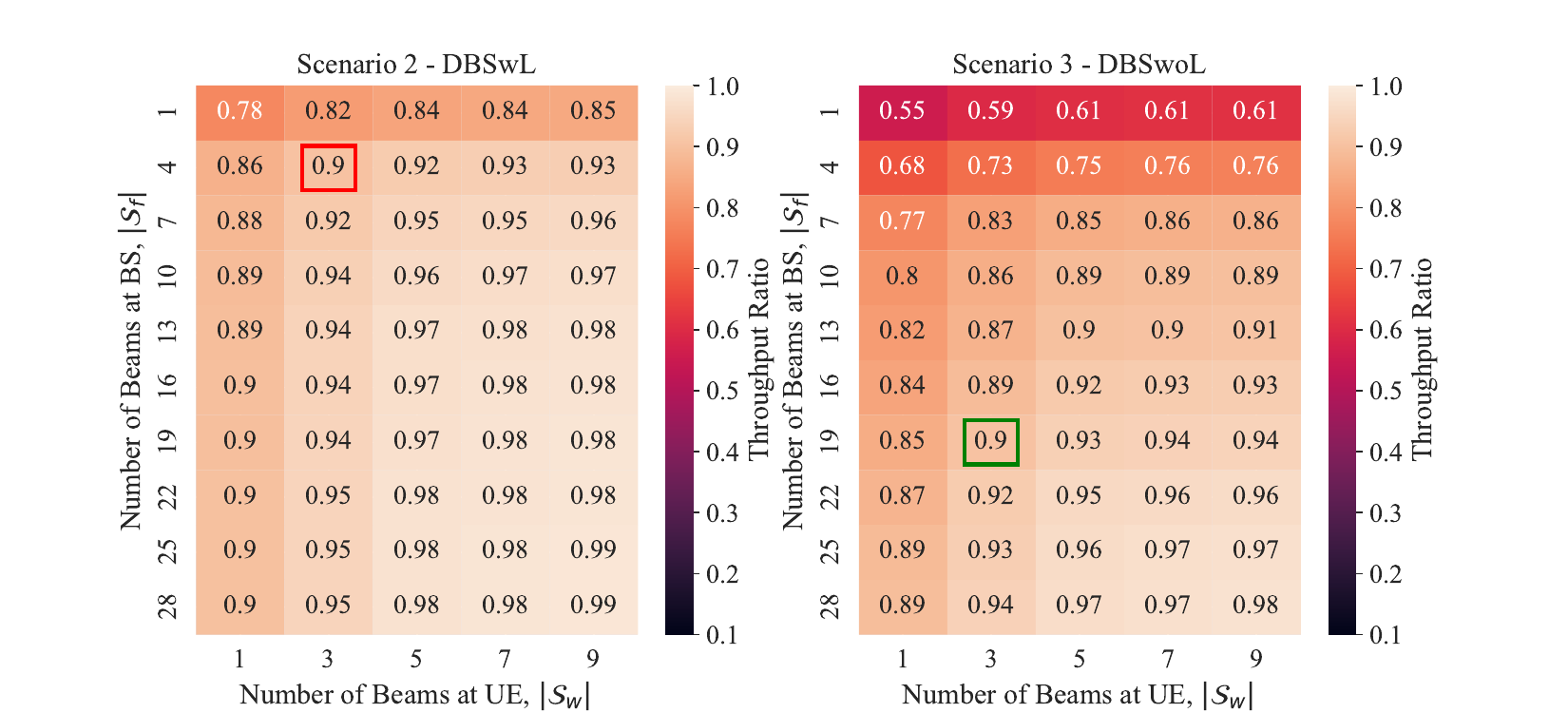}
	}
	\caption{Throughput ratios for \textit{scenario 2} and \textit{scenario 3}. The horizontal and vertical axes represent the number of the selected UE and BS beams. Heatmaps are useful to identify the required number of BS and UE beams to achieve a certain performance. For example, $4$ BS beams and $3$ UE beams are sufficient to achieve $90\%$ throughput ratio in {\scenabbtwo} whereas {\scenabbthree} requires $19$ BS beams and $3$ UE beams, which is equivalent to say $45$ more beam pairs are required compared to {\scenabbtwo}.}
	\label{fig:s3_s2_comparison}
\end{figure}

\figref{fig:s3_s2_comparison} illustrates the throughput ratio comparison between \textit{scenario 2} and \textit{scenario 3}. It shows a significant increase in throughput ratios from $\card{\cS_\text{w}} = 1$ to $5$, with less notable growth from $\card{\cS_\text{w}} = 5$ to $7$. This shows that $5$ beams from the UEs' codebook are enough to achieve approximately $95\%$ of achievable throughput ratio as seen in \figref{fig:throughput_ratio_all} and \figref{fig:s3_s2_comparison}. Beam selection at the BS without the UE's location causes a throughput ratio decrease of over $20\%$ in single BS beam selection. Selecting multiple BS beams leads to a notable performance increase in  \textit{scenario 3}. This shows the effectiveness of our clustering approach in leveraging location information from the dataset to select beamformers, even without the UE's location. 
Moreover, \textit{scenario 3}  performs almost the same as \textit{scenario 2} after $100$ beam pairs. As the number of clusters increases, the clustering approach recovers the performance loss as if the UE's location is available at the BS.

 \begin{figure}
    \centerline{
     \includegraphics[width=\figsizefive \textwidth,trim={0 0 0 36},clip]{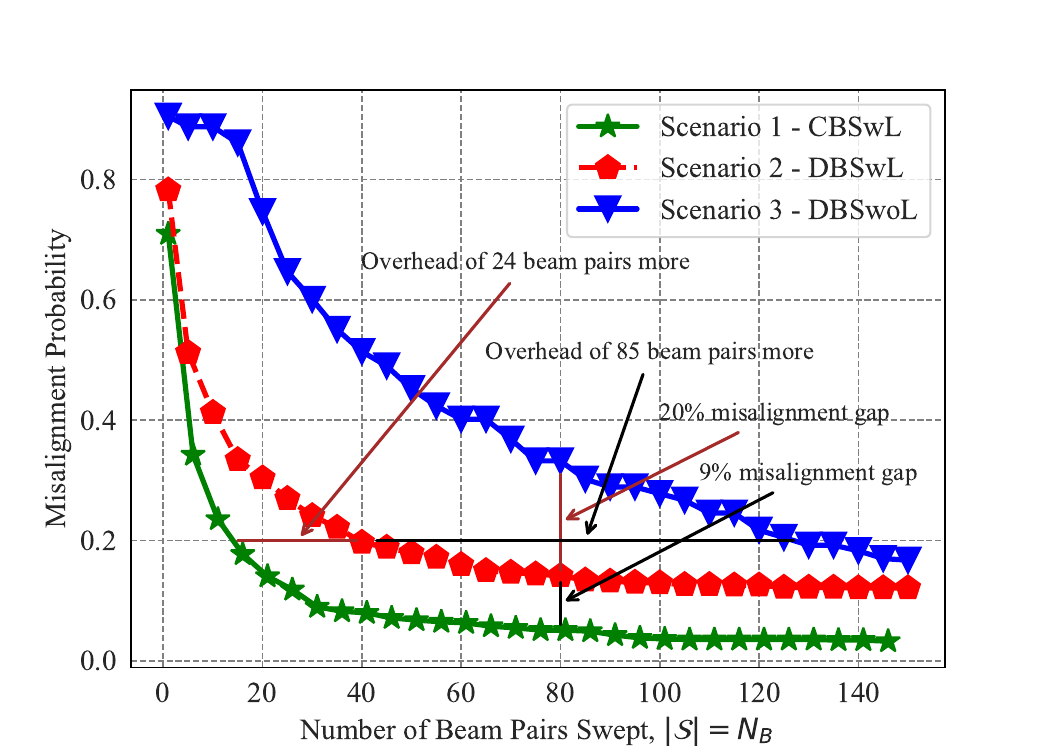}
    }
    \caption{Misalignment probability comparison for the three scenarios. The number of UE beams, $\card{\cS_\text{w}}$ is set to 5 for the decoupled scenarios. There is a large gap between the coupled \textit{scenario 1} and decoupled \textit{scenario 2} and \textit{3}. The decoupled scenarios might not necessarily yield the best beam pair since they are designed to achieve a high throughput ratio, not the best beam pair.}
    \label{fig:powerloss_all}
\end{figure}

\figref{fig:powerloss_all} demonstrates the misalignment probability for the three scenarios. \textit{Scenario 1} achieves significantly lower misalignment probability than decoupled scenarios for any number of beam pair selections. \ul{Even though there is a huge performance loss in misalignment probability, not selecting the best beam pair does not lead to a significant performance loss in the throughput ratio}. This shows there are other suboptimal beam pairs that still yield a high throughput ratio which is what our ML methods are designed to achieve. \figref{fig:throughput_ratio_all} shows that the throughput ratio is still near $100\%$ when the misalignment probability is high. \textit{Scenario 2}  and \textit{scenario 3}  perform quite well in terms of throughput ratio even though they are not able to include the best beam pair with high probability.

\section{Conclusion and future work}

In this work, we analyzed the impact of decoupling beam pair selection and disaggregating location information from the BS on beam training in dynamic V2I mmWave MIMO communication systems. We proposed location-aided ML-based beam selection methods for decoupling. The results showed that ML-based decoupled beam selection with location information has almost no performance decrease in throughput ratio compared to the conventional coupled beam pair selection with the location. 
Moreover, the proposed clustering-based beam selection approach gradually recovers the performance loss, although disaggregating location has a noticeable performance loss compared to decoupling beam selection. In summary, our work provides promising results for decoupling training between the BS and the UE with or without localization. In future work, we will expand on this work by analyzing how to address the practical aspects of decoupled beam training and how to exploit decoupling for heterogeneous devices. 

\section{Acknowledgement}
This material is based upon work supported by the National Science Foundation under Grant $\text{NSF-ECCS-2153698}$, Grant $\text{NSF-CCF-2225555}$, Grant $\text{NSF-CNS-2147955}$ and is supported in part by funds from federal agency and industry partners as specified in the Resilient \& Intelligent Next Generation Systems (RINGS) program.

\bibliographystyle{IEEEtran} 
\bibliography{./references.bib}

\end{document}